\documentclass[conference]{IEEEtran}
 \usepackage{mathptmx}
 \usepackage{graphicx}
 \usepackage{times}
 \usepackage{balance} 

\usepackage{hyperref}

\usepackage{enumitem}
\setlist{nolistsep}

\title{Making ethical decisions for the immersive web}

\author{Diane Hosfelt \\ dhosfelt@mozilla.com \\ Mozilla \\ }
\date{\today}

\begin{document}

\maketitle

\begin{abstract}
Mixed reality (MR) ethics occupies a space that intersects with web ethics, emerging tech ethics, healthcare ethics and product ethics (among others). This paper focuses on how we can build an immersive web that encourages ethical development and usage. The technology is beyond emerging (footnote: generally, the ethics of emerging technologies are focused on ethical assessments of research and innovation), but not quite entrenched. We're still in a position to intervene in the development process, instead of attempting to retrofit ethical decisions into an established design. While we have a wider range of data to analyze than most emerging technologies, we're still in a much more speculative state than entrenched technologies. This space is a challenge and an opportunity.
\end{abstract}

\section{Introduction}
With the advent of affordable devices like Oculus Go and AR-enabled smartphones, mixed reality devices are hitting the mainstream market. Market growth projections over the next five years range from 40-80\%. As we've experienced with the ubiquity of artificial intelligence (AI) and machine learning (ML), the development of ethical frameworks and guidelines tends to lag behind the technology itself. Once we begin to fully recognize the impact of new technologies, we're left to retrofit regulations and ethical decision making into technology that's already had billions of dollars invested in it.


Mixed reality (MR) devices blend digital elements and the physical world, covering a wide spectrum that includes both virtual reality (VR) and augmented reality (AR). In VR, a device occludes the user's vision (and often other senses) to present a fully digital experience, while AR experiences overlay digital elements on users' perceptions of the physical world. Two other terms that often appear when discussing MR are "spatial computing" or "immersive technologies."


The contributions of this paper are to outline the unique risks posed by MR devices, present plausible scenarios that may result from insufficient protections, and propose steps that will improve protections from legal, regulatory, societal, and engineering standpoints.

\section{What are Mixed Reality Ethics?}
Tech companies have embraced the mantras of 'ask for forgiveness not permission' and 'move fast and break things' to the detriment of individuals' privacy, security, and safety. Instead of thoughtfully approaching difficult problems and considering how we can prevent abuse, we sell applications that actively aid abusers~\cite{dell}. Instead of designing systems to empower and protect users, we create environments that foster harassment without clear or sufficient accountability mechanisms~\cite{dreyfuss}. Instead of debiasing algorithms and training data, we build self-driving cars that are more likely to hit dark skinned people~\cite{wilson2019predictive}.

By building technology that violates users' privacy and denies them agency, we're creating a dystopian future. Mixed reality, with its ability to combine virtual and physical elements, is a powerful mechanism for distorting our perspectives.  In 1987, Simitis argued that large scale data collection is "the ideal means to adapt an individual to a predetermined, standardized behavior that aims at the highest possible degree of compliance with the model patient, consumer, taxpayer, employee, or citizen."~\cite{simitis} If we extend this reasoning to the scale of data collection today and then consider incorporating MR-derived data, it's clear that we need to embrace ethical principles before it's too late.

Generally, the ethics of emerging technologies are focused on ethical assessments of research and innovation. Mixed reality (MR) ethics occupies a space that encompasses emerging tech ethics, healthcare ethics and product ethics. The technology is beyond emerging, but not quite entrenched. We're still in a position to intervene in the development process, instead of attempting to retrofit ethical decisions into an established design. 

This paper focuses on issues that face building a platform that encourages ethical development and usage. While we have a wider range of data to analyze than most emerging technologies, we're still in a much more speculative state than entrenched technologies. This space is a challenge and an opportunity.

\subsection{Web ethics}

For our purposes, there's also an overlap with web ethics, because we're building a platform for the immersive web. Having an open and accessible web means that we can invite more diverse viewpoints. Today, if you want to be an iOS or Android app author, you must be in an approved nation; otherwise, you and your country can't participate in the ecosystem.

The web is shaped by standards bodies like the World Wide Web Consortium (W3C). Standards are crafted by consensus, so the intentions of a single bad actor are minimized.

When one browser has a monopoly, developers aren't incentivized to make their sites work on multiple platforms. This decreases both competition and the efficacy of web standards---if there's only one major browser, then their implementation becomes synonymous with the standard. We need diverse viewpoints and interests to shape the web, otherwise it will only serve the interests of a few.

The immersive web has a number of advantages over a solely app-based ecosystem. Unlike apps, there are no inherent restrictions on who can develop or access web resources. It's also intended to be cross-platform, allowing users with a \$300 MR device to have a similar experience to those with a \$3000 device. Perhaps most importantly, it allows web browsers to act as a trusted intermediary for device resource requests. Instead of a native app running in the background with access to information like orientation data, the webpage needs to request this through the browser, which could reject inappropriate requests.

\subsection{Mixed reality ethics}

Mixed reality technology has the potential to transform the way we interact with each other and the world around us. The best way to level out asymmetries of knowledge and power is to not allow them to form in the first place. This paper is an initial exploration into the challenges we face while we try to define what an ethical immersive future looks like. I propose the following principles of building ethical software in mixed reality:

\begin{itemize}
	\item Ask permission, not forgiveness
	\item Minimize tracking and fingerprinting via biometrics
	\item Empower individuals to define how they're perceived virtually
	\item Prioritize mechanisms for reporting harassment and blocking perpetrators
	\item Identify ways to incentivize the principle of least privilege
	\item Consider privacy a first-class requirement
	\item Be both transparent and accountable
\end{itemize}

Section \ref{sec:laws} briefly discusses relevant laws, and section \ref{sec:steps} proposes concrete steps we can take to embrace that these principles.

\section{Data collection and inference}\label{sec:data}
Why are immersive ethics any different from other technologies? Immersive technologies, whether augmented or virtual, affect our physical bodies in ways that non-immersive technologies don't. Head-mounted displays (HMDs) overlay and mix virtual elements on our senses, changing our perceptions of ourselves and our surroundings. Sometimes, this can even result in 'cybersickness.'

What about non-immersive, hand-held, AR? While it doesn't have the same physical effects on users, it shares many of the same data privacy concerns. MR experiences continuously collect and process environmental data in near-real time. In this case, the data available is much more extensive than even the intrusive data collection we've become accustomed to currently.

Spatial computing and immersive experiences expose, by necessity, information that poses a threat to privacy. To enable these technologies, we rely on many extended duration sensors. These sensors fall into three categories: biometric, orientation, and environmental.

\subsection{Biometric data}\label{sec:biometrics}
A wide range of biometrics can be collected by head-mounted displays (HMDs), some of which are non-obvious to users. In addition to eye-tracking data, we can collect information on users' gait, height, and physical/emotional reactions.

Biometric information presents particularly difficult problems. Firstly, once exposed, there's no way to retrieve or change it. Even worse, it provides methods for fingerprinting users by their physical attributes, not just their online behaviors. Biometrics also provide insight into involuntary nonverbal reactions\cite{bailenson2018protecting}. Pupil dilation and skin temperature can indicate a user's sexual attraction or orientation. Gaze tracking can expose details of medical conditions like autism or anxiety disorders. Innocuous data like facial movements during a task can classify people as high or low performers~\cite{jabon2011automatically}.

\subsubsection{Scenario: Nonverbal data and job interviews}
Some companies, like Unilever, are currently deploying emotion detection technology to predict how job applicants will react to certain situations~\cite{gilliland}. Others, like Lloyd's Bank, are putting applicants in VR simulations for similar purposes~\cite{guardian2018how}. While immersive technologies can improve geographic restrictions on interviewing or working by allowing virtual colocation, there are negative implications. Consider an applicant who is interviewing virtually for a position at a company led by a CEO whose personal religious beliefs maintain that homosexuality is immoral. During the interview, the headset detects nonverbal reactions from the candidate that suggest they may be gay, and the company's algorithms (possibly opaque to the interviewer) reject the applicant.

Wouldn't this be illegal? In some countries, maybe. However, if the algorithm is simply trained to reject certain behaviors, not to explicitly exclude certain sexual orientations, it might not be. After all, the interviewer didn't ask about the applicant's sexuality. The technology just detected that their personality isn't 'suitable,' whether or not the applicant is gay or not.

This scenario highlights the broad intersections of tech ethics in the MR space. First, we have AI ethics---is it acceptable to train algorithms that reject job applicants? Should such algorithms output details on the behaviors detected and decisions made (and will a human be able to understand the details)? Should we be creating algorithms that can identify sexual orientations? MR's unique contribution to this situation is the sheer amount of nonverbal data it collects in short periods of time.

This data will be misused and the consequences could be life-altering. If we don't take action now on the privacy issues presented by nonverbal data, we'll either abandon promising technology altogether, or live in a dystopia.

\subsection{World data}

Particularly for AR applications, we need to incorporate world data to the virtual model. While data about the physical world is collected using cameras, devices don't have to provide all information to the application. Instead of providing full camera access, platforms can provide limited hit testing or a world mesh.

We need to identify ways to incentivize the principle of least privilege---it's easy to provide full camera access to applications and let them figure out what they want to use. However, that shouldn't be the default option. World meshes and hit testing provide sufficient spatial data for many applications without also transmitting details like text.

\subsubsection{Scenario: Camera access and health data}\label{sec:data:world:camera}

We know that advertisers are interested in users' health data~\cite{jeong2019insurers}. Consider the classic AR example: an interior design application that places virtual furniture in your home. I often leave my medications on my nightstand, so that I remember to take them before bed. It's plausible that when I'm redesigning my bedroom, the application will detect the medication, identify it (either by the unique pill shape or by detecting and reading the label), then transmit this information to third parties, which will then use this information to target me for ads related to my condition.

There are a number of reasons AR apps tend to default to full camera access. Sometimes, the libraries applications depend on require more permissions than the application actually uses; however, by using the library, they must request enhanced sensor access. Most importantly, it's easier. Applications can determine what information they need and discard the remainder...or they can take the surplus data and turn it into a new revenue stream, a lucrative practice pioneered by Google~\cite{zuboff2019age}.

Google initially ignored the collateral data produced by search queries until engineers realized that this 'data exhaust' could be used to model users' behavior. At first, this operated as a 'behavioral value reinvestment cycle,' where Google harvested user data to improve the search product. Later on, engineers realized that this behavioral data surplus could also be used to create detailed user profiles and target ads more successfully~\cite{patent2003targetedad}.

There are numerous ethical concerns with this scenario. Is it acceptable for advertisers to target users based on medical data? Is it acceptable for applications to gather data like this, unrelated to the purpose of the application?  A legitimate use of the same data that is problematic in this instance would be an application that identifies pills and their uses, possibly as an aid for healthcare professionals.

As a platform for creating immersive applications, we're faced with a complex problem: how can we enable legitimate uses of sensitive data while discouraging misuse? One potential way to approach this specific problem is to recognize that medications and prescription labels are often highly standardized. We could use object recognition techniques to detect the labels, then not provide that information to the application unless the user explicitly grants further permissions.

\subsection{Orientation data}

MR devices use accelerometers, magnetometers, and gyroscopes to orient themselves in the physical world. Because of permission fatigue, devices don't ask permission for all sensor accesses. Instead, they sort sensors into two categories: dangerous and not. 'Dangerous' sensors include microphones, cameras, and GPS, while orientation sensors are considered 'not dangerous.'

However, it turns out that 'not dangerous' sensors also pose serious concerns to user privacy. For example, we can use the accelerometer or ambient light sensor of a cellphone to extract the user's screen lock pin~\cite{aviv2012practicality, spreitzer2018systematic}. The interactions between sensors pose a threat to our existing security models. It's difficult to anticipate the potential side channel attacks that existing sensors pose, let alone predict how additional sensors may create novel threat vectors.

Permissions have already become too complex to easily communicate to users what data is gathered and the potential consequences of its use or misuse.

\section{Applications and ethical scenarios}
\subsection{Communications medium and social spaces} \label{subsec:communications}
The immersive web gives us new ways to connect and represent ourselves. In an instant, you can be 'present' somewhere on the other side of the world. It's the closest we've come to apparition or teleportation.

There's been an explosion of social VR platforms---AltspaceVR, VRchat, Facebook Spaces, Rec Room, Mozilla Hubs, and Anyland to name a few. Each takes a different approach to moderation and governance. They all have some commonalities---avatars and interactions. Maloney identifies three main ethical considerations for avatars~\cite{maloney}:
\begin{description}
	\item[ Effects of perceptual manipulations]: Immersive experiences can violate physical laws and manipulate or deform body parts. How will amputees react to having four limbs in VR, but not in the physical world? Do we need to recalibrate users to the limitations of the physical world after certain VR experiences? How can we ethically study causes and prevention of cybersickness?
	\item [Negative effects caused by your avatar]: Avatar choice can effect our self perceptions even after exiting experiences. Users have experienced increased self-objectification after embodying sexualized avatars and self-imposed stereotypes. However, avatars can also affect positive behaviors---rendering users as their "future selves" can lead to increased saving behavior. How do we balance these manipulations with informing users?
	\item [Negative effects caused by others' avatars]: Representations of avatars can lead to negative stereotype confirmation, and users are less likely to collaborate with avatars that represent diverse ethnic groups. How can platforms and communities balance self-expression while preventing hate speech and minimizing bias?
\end{description}

While harassment may fall into the category of 'negative effects caused by others' avatars,' this would be too limiting. Due to the unique nature of social VR, harassers can combine the anonymity and capability of other internet social spaces (e.g. threatening text messages, inappropriate verbal conduct) with the avatar's presence to grope or make obscene gestures. In a study of 609 VR users, 36\% of men and 49\% of women experienced sexual harassment in VR~\cite{outlaw2018}.

Unfortunately, social VR enables physical harassment to occur regardless of physical distance, because the VR-enabled embodiment  makes harassment more intense~\cite{blackwell}.

Defining harassment and providing reporting mechanisms is ongoing, particularly since definitions of harassment are subjective. Outlaw found that the most effective tools for dealing with harassment were blocking and muting harassers. In a separate study focused on women in VR spaces, all respondents reported feeling unsafe and uncomfortable after spending 30 minutes in social VR and went out of their way to avoid attention~\cite{outlaw2017}.

MR devices can also be used to enrich a user's information about the physical world. For example, consider a headset that enriches the user's worldview with sentiment data. The headset could detect facial expressions and more subtle nonverbal cues, like pupil dilation, to determine bystanders' mood and reactions. How is this different than going into a public space and simply looking around, inferring the emotions of those around you? A key differentiator is scale---a computer could analyze the emotions of everyone in the field of view while simultaneously integrating cues that wouldn't be detectable to most humans.

The bystanders haven't had the opportunity to consent to this type of analysis. While they are in a public space, that doesn't mean that shouldn't have some expectation of privacy (namely, the right to not have their face and body recorded and analyzed)? In the US, you have 'reasonable expectation of privacy' in a public space (consider the "plain view" doctrine); however, the emergence of always on cameras and microphones suggests that we may need to reevaluate what privacy is 'reasonable' given the rapid technological advances and prevalence of such devices.

The same scenario can also illustrate a different set of ethical concerns. Suppose we live in a world where HMDs are common everyday wear. Instead of relying on my device to interpret others' emotions, their devices would communicate this information to mine. In this instance, perhaps they've consented to allowing their device to collect that information on them, but does that mean the device should be allowed to transmit it?

Instead of using nonverbal cues to infer emotions, the device could use facial recognition to remind user's of a person's name and everything they should remember about the person (names of spouse and children, how you met, last topic of conversation, etc.) to avoid embarrassment at cocktail parties\cite{wassom2014augmented}? Is this less invasive? Does the private setting of a party change the ethical considerations?


\subsection{Privacy}
Augmented and virtual reality create, and can only really exist, in a world where cameras and computer vision applied to their outputs are more or less ubiquitous. They require data sources that can reveal individuals' intrinsic characteristics and behaviors (e.g. pose, movement, head tracking) for minimal functionality. Defining and defending privacy in this world is an existential question for the technology.

Section \ref{sec:data} discusses some of the implications of the data MR devices collect and process in order to function. Privacy is difficult to define, because we each have different risks and threats to evaluate. A one-size-fits-all approach won't suffice. While this is a pervasive issue in tech today, MR introduces new opportunities and considerations for violating users' privacy.

It has become common to require that people pay-for-privacy. Consider shopping: if you share your address (physical or email), then you can enjoy increased discounts. Nearly 1 in 4 respondents to a holiday shopping survey said they would not hesitate to provide their personal information for better deals~\cite{moses}. The trouble is that there's so much data now, that we aren't just giving an address for better deals. We're enabling companies to build and share huge user profiles. Facebook distributed an application, called Facebook Research, that paid users \$20 in giftcards per month to allow Facebook nearly unfettered access to their devices~\cite{axon}.

A notable example is the Washington Post. After the European Union passed the General Data Protection Regulation (GDPR), the Washington Post responded with two different subscriptions---\$90 for the GDPR-compliant EU version, \$60 for the US version~\cite{karl}.

\subsubsection{Scenario: Facebook Research, Oculus edition}

What could Facebook Research look like for HMDs? Facebook's Oculus devices are becoming more affordable for the masses?consider an initiative that provides free devices to low-income students. These devices offer access to learning opportunities and advanced courses that aren't available in the student's local school. In exchange, they (and their parents) consent to providing Facebook with all data collected by the device. The resulting data would create a comprehensive profile of the child?their physical development, voice, educational progress, emotional maturity, etc.

In pursuing better education at a price they can afford, the child has instead paid with their current and future privacy. We can anticipate a number of possible scenarios that could follow:

Suppose that one day, the student's gait changes. Facebook's algorithms identify that the child has likely sprained their ankle and serves ads for crutches, orthopedic doctors, homeopathic cures, etc.

Now, suppose that in high school, the student struggles for a period---skipping class, getting into fights, poor grades---due to difficulties at home. Although the student's behavior and grades improve, they're labeled as a troublemaker or a low performer. Later on, they struggle to get a job, because companies pay Facebook for additional information on applicants.

\subsubsection{Scenario: Schools, VR and neuropsychological diagnosis}

Schools provide a vital, though controversial, role in children's health. In addition to teachers and nurses administering necessary medication, teachers often act as "disease-spotters" for disorders like ADHD~\cite{phillips2006medicine}. If a teacher suspects that a student may have a disorder, they would alert the parents, who would then need to have a doctor diagnose ADHD. Diagnosis is difficult, but research has shown that using VR can improve neuropsychological assessments~\cite{areces2018analysis}. As VR becomes widely deployed in classrooms, it may be tempting for schools to monitor children's interactions in virtual environments to assess whether they have these conditions.

What happens if we have schools collecting this type of data on students? Is it ethical for them to interfere in students' health like this? Will they create student profiles that indicate 'ADHD tendencies' without a formal diagnosis? How will educators treat them differently? Will it follow them throughout their entire education?

\subsection{Accessibility and inclusion}

The immersive web gives us new ways to connect and represent ourselves. We need to design systems to prevent abuse and harassment as first class requirements while empowering users to choose how they are represented and recognized on the immersive web.

Current HMDs largely rely on motion controls and require users to take certain positions (i.e. standing). While we can create accessible applications on an immersive web, they're not usable if the HMD can't accommodate them. As a society, we've recognized that excluding people based on disability is as unethical as excluding them based on skin color, gender identity, or sexual orientation.

MR devices need to integrate appropriate accommodations, like controllers that provide haptic feedback~\cite{zhao2018demonstration} or settings that allow users to indicate that they're in a wheelchair so that the device doesn't keep insisting that they stand.

MR also has unique potential as an assistive technology. VR presence and embodiment can allow wheelchair bound users to ski~\cite{harrell} or allow elderly relatives to participate in family events (even if they can't travel). For visually impaired users, MR could read signs or papers out loud. It can also help with navigation by reading out directions in real time and detecting oncoming traffic. For hearing impaired people, MR devices can provide personal subtitles in theaters~\cite{forrest} and translate public announcements (like train announcements) into text in real time. They could also interpret group conversations into subtitles while providing speaker attribution.

Unfortunately, many accessibility features also pose privacy concerns.

\subsubsection{Scenario: Real-time 'subtitles'}

For people who are deaf or hearing-impaired, group conversations can be particularly difficult to follow. Speech processing has made it possible to have real-time captions~\cite{welch}, which could then be displayed on the user's HMD. In a group conversation, this would likely be jumbled; however, it's plausible that we can use the spatial characteristics of audio to determine who says what. Then, the captions could be displayed to indicate who said what, allowing the user to more easily follow the conversation.

What ethical concerns might exist in this situation? Is this kind of video and audio processing considered recording? Do the bystanders need to consent to this? If the data is sent to the cloud for processing, do the participants have a reasonable expectation of privacy? What happens if the device inadvertently transcribes a conversation the user wasn't intended to hear?

\subsubsection{Scenario: Gaze-based navigation}

People with mobility impairments are excluded from using technology that's largely touch centric, like phones and tablets. Workarounds exist, like voice control and dedicated 'accessible apps,' but tasks that are often considered simple, like navigating a web page or answering a call, become more complex. What if we could use gaze instead? This would allow people with very limited mobility to fully navigate an immersive world. User input in MR is already largely speech based, because text input is difficult and time-consuming in HMDs.

As mentioned in section \ref{sec:biometrics}, gaze tracking is a powerful biometric. Not only can our eyes indicate sexual attraction and other nonverbal reactions, but they can reveal details about our decision making process~\cite{costandi}.

Is it ethical to require disabled users to sacrifice that much privacy in order to use modern technology?

\subsubsection{Representation}

MR should be a tool for inclusion and representation; for example, immunocompromised students could be 'present' in the classroom without risking their health.  In addition to accommodating disabilities, HMDs also need to fit over different hair styles and track different skin colors. As MR becomes more prevalent in education, the problems of excluding individuals because the HMDs don't fit properly become even more obvious.

As social MR becomes more prevalent, we need to build platforms that are flexible enough to allow individuals to shape their experiences and how they're perceived. For example, offering two gender options for avatars would prevent a nonbinary person from being adequately represented. Do avatars even need to be human? Should humanoid avatars accurately represent the physical attributes of a person? We need to balance giving individuals the option to accurately represent themselves (by providing adequate skin/hair/body options) while accepting that they may choose a different appearance. For example, a woman may choose a male avatar when entering a virtual space to avoid harassment.

Unfortunately, just like in the physical world, individuals will choose to express themselves in ways others may consider inappropriate or obscene. Suppose a user creates an avatar that displays a Nazi symbol. Hopefully, the platform would ban any displays of hate symbols, and provide easy mechanisms for reporting and banning anyone who violates that. While this might not be illegal in the US, other countries, like Germany, have laws prohibiting Nazi symbols and other forms of hate speech---does the platform have a duty to cooperate? While we develop technologies that allow us to be 'present' anywhere in the world, we need to consider how international laws and norms impact our platforms.

\section{Discussion}
Immersive technologies occupy a space that intersects emerging, enabling, and entrenched technologies. While we have some concrete data and examples of misuse, we're also in a speculative position---what is the potential and how can it be abused? As a platform, the immersive web has less specific considerations than stand-alone technologies that will be built on top of it. For example, an education app may ask "how can we create a virtual classroom without violating students' privacy." They would plausibly consider minimizing collection of biometric data that could inadvertently reveal health conditions and whether it's acceptable to infer that a student's attention has wandered and manipulate them back to the task at hand. As a platform, we need to consider how we build thoughtful controls for accessing the biometric data that could enable this.

This means that we also need to understand the current legal concept of privacy and how it might apply to future technologies. By identifying gaps at this point in time, we can collaborate with legislators to craft regulations that are useful and technically feasible. For example, instituting a blanket ban on collecting biometric data would be less useful than preventing companies from selling information to third parties that's been derived from the biometric data.

\subsection{Current legal landscape}\label{sec:laws}

This section focuses on the existing information privacy law landscape and how we might apply it to the challenges presented by MR devices. With the explosion of technical advances that have occurred over the past few decades, it's not surprising that the law has struggled to keep pace. In his \emph{Olmstead v United States} dissent, Justice Louis Brandeis states that "[the government] relies on the language of the [Fourth] amendment, and it claims that the protection given thereby cannot properly be held to include a telephone conversation... this Court has repeatedly sustained the exercise of power by Congress, under various clauses of that instrument, over objects of which the Fathers could not have dreamed."

Privacy isn't just about what the law says---it's shaped by the society in which we live and how we value privacy. Ethics and society can inform the law, and in turn, the law should reinforce ethical and social norms. While some aspects of MR devices are inadequately addressed by current laws, there are some clear parallels we can make.

Consider the 'emotion-annotating'  HMD presented in \autoref{subsec:communications} and its resemblance to polygraph machines. Modern polygraphs use similar sensors to those used in MR devices, including GSR, blood pressure, and respiration, to detect if subjects are being untruthful. However, most courts exclude polygraphs as evidence, citing weak scientific underpinnings. Despite this, many employers still attempt to use them during hiring or investigations. Thanks to the Employee Polygraph Protection Act, private sector employers are disallowed from requiring polygraphs or disciplining them solely on the basis of the results of a polygraph test, but there are some circumstances where it is lawful to use them.

There are limits on what questions can be asked by polygraph examiners, namely questions regarding religion, race, politics, sexual behavior, or union activities. Based on this existing precedent, we can say that the law \emph{should} similarly prevent the use of MR devices to infer those beliefs. Unfortunately, it may be more difficult to prove that an MR device inferred this information than it is to show that a polygraph examiner asked a question illegally.

The US Constitution doesn't explicitly state a right to privacy, but the courts have asserted that it can be inferred by the First, Third, Fourth, and Fifth amendments. However, this only applies to intrusions by the government; instead, private sector data collection and use are regulated by a patchwork of federal and state laws. Instead of blanket privacy laws, sector-specific privacy laws are more common, such as the Video Privacy Protection Act and the Children's Online Privacy Protection Act.

\begin{quote}
The sectoral approach in the United States can sometimes draw even finer distinctions for similar kinds of information. For example, cable TV records are regulated differently from video rental or sale records. There are no industry specific federal statutes directed towards the personal information contained in the records of most merchants.\cite{solove2018information}
\end{quote} 

\subsubsection{Personally identifiable information (PII)}
The concept of PII is central to most privacy legislation, so, what is it? We don't actually have a good definition.
\begin{description}
	\item[Tautological]: "PII is any information that identifies a person"
	\item[Non-public]: PII is non-public information (but doesn't mention identifiability)
	\item[Specific types]: lists data that is PII, such as name and address
\end{description}

I contend that the tautological approach is the least bad of these, because it is more adaptable to new technologies. For example, suppose an individual's movements in a virtual world create a unique, identifying fingerprint. Would that be PII? It wouldn't be listed as PII in legislation, and it's not necessarily 'non-public,' but it is an identifying biometric that we have an interest in protecting.

How do we address and protect PII in an immersive world? How will we even define PII?

\subsubsection{Reasonable expectation of privacy}
The "reasonable expectation of privacy" test was established in a concurring opinion for \emph{Katz v United States}\cite{katz1967} to articulate when the Fourth Amendment applies:
\begin{enumerate}
	\item A person must exhibit an "actual (subjective) expectation of privacy"
	\item "The expectation [must] be one that society is  prepared to recognize as 'reasonable' "
\end{enumerate}

The Court ruled that it was a violation for the police to record Katz's conversations in a phone booth, noting that "the Fourth Amendment protects people, not places." When we look at applying these principles to MR technology and experiences, we don't have well-defined concept of what privacy is in the space. As a society, we need to determine what 'reasonable privacy' is in virtual spaces. This is complicated by the \emph{Third Party Doctrine} established by \emph{Smith v Maryland}\cite{1979smith}, which holds that individuals have no reasonable expectation of privacy over data they've 'voluntarily' given to third parties.

\begin{quote}
Marshall, J. joined by Brennan, J. dissenting...Privacy is not a discrete commodity, possessed absolutely or not at all...In so ruling, the Court determines that individuals who convey information to third parties have 'assumed the risk' of disclosure to the government. 
\end{quote}

The Third Party Doctrine hasn't adapted well to the digital age---should we be forced to forgo privacy because companies collect massive amounts of personal and behavioral data on us? Justice Sonia Sotomayor argues this in her concurring opinion for \emph{United States v Jones}\cite{jones2012}. Solove contends that maintaining the Third Part Doctrine in a world where businesses maintain detailed digital dossiers will violate individual's First Amendment rights and leave them vulnerable to government abuses\cite{solove2001digital}. This also raises concerns with the concept that individuals are voluntarily providing data to third parties. Is it possible to make a phone call without providing the number to your provider? Or to browse the internet without giving data to your ISP? These are common ways we participate in society; it's unethical to require that individuals avoid them to preserve their privacy.

Consider the role of this doctrine in the immersive age. Companies will have to collect information on our gaze, motion, and physical responses to enable certain MR experiences. Are we prepared to give up the ability to make decisions without constantly worrying that the government is monitoring our eye-tracking data (revealing internal thought processes)? Instead, shouldn't companies be forbidden from misusing or disclosing this data? 


\subsubsection{Behavioral data and marketing}

Targeted marketing often avoids the privacy problem by asserting that the behavioral data they use isn't PII and users are anonymized. However, this ignores the power of technology to deanonymize such information \cite{narayanan2008robust}. In \emph{NAACP v. Alabama}\cite{1958naacp}, the Court ruled that compelling the NAACP to disclose information about members would violate their right to freedom of association, because it depends on their privacy. If their membership was revealed, this could lead to ostracism. MR provides even richer behavioral and biometric data sources. We haven't yet explored deanonymization in the space, but we expect it will be possible.

\subsubsection{Privacy Policies}

Most companies post privacy policies on their webpages. In these, it's common for companies to state that they can use and disclose personal information however they'd like unless the consumer opts out, emphasizing the role of individual choice in US privacy management. However, Daniel Solove "contend[s] that [privacy self-management] is being tasked with doing work beyond its capabilities....[and] does not provide people with meaningful control over their data' \cite{solove2012introduction}. In particular, the use of big data analytics precludes non-experts from understanding many privacy implications \cite{baruh2017big}. This isn't a failure of individuals' comprehension, but rather a deliberate barrier---companies have a vested interest in maintaining their access to behavioral marketing and selling data to third parties.

In addition to informing users (although, in practice, consumers rarely read privacy policies), privacy policies have become a way to make companies accountable. The Federal Trade Commission (FTC) considers violating privacy policies to be an unfair and deceptive practice. This has led to the FTC being the leading privacy agency in the US, despite being limited in enforcement capabilities.

\subsubsection{International Privacy Law}

American laws are focused on protecting the individual's domain (particularly their home) from the state, while European laws treat privacy as fundamental part of human dignity.
In contrast to the US's sectoral approach to privacy, Europe has crafted comprehensive information privacy laws. The European Union's new data privacy law, the General Data Protection Regulation (GDPR) creates strong legal protections for individual rights, limits processing and collection of sensitive data, and increases enforcement tools, like fines. A key element of the GDPR is disallowing individuals from 'opting out' of the fundamental protections to prevent companies from gathering and processing data that is beyond the purpose of the contract.

The complexities introduced by requiring companies to comply with different privacy laws across the world (some of which are directly opposed to others) will only get worse as immersive technologies make the world even smaller.

\subsection{Concrete steps for ethical decision making}\label{sec:steps}
\subsubsection{Educate and assist lawmakers}
If technologists don't coordinate with legislators, we'll end up with rigid regulations that create a lot of paperwork, but end up being privacy theater. This is exacerbated in the US, where states have been leading the charge for better privacy legislation, forcing companies to comply with a hodgepodge of laws.

By working with legislators, we can help them understand the complications and unique considerations for mixed reality technology. We can also try to craft legislation that won't create an undue burden on smaller companies---companies spent millions of dollars retrofitting their systems for GDPR compliance. Some companies chose not to comply and stopped serving European users to avoid penalties.

\subsubsection{Establish a regulatory authority for flexible and responsive oversight}

It's unrealistic to expect laws to keep up with the pace of technological advancement. For bills signed into law from 2011-2019, it took 241 days between for a bill to become a law. Generally, this is a good thing. Stable laws create a stable government. Unfortunately, this means that the laws are significantly behind, and they're written and interpreted by people who don't have much technical background.

The Financial Industry Regulatory Authority (FINRA) was established in 2017 to be a self-regulatory organization (SRO) for the finance industry. Its purpose is to promote trust in the US finance industry by ensuring that firms operate fairly. The tech industry can similarly embrace an SRO to create ethical guidelines and craft (and enforce) regulation to better serve consumers\cite{reich}.

Like the model of web governance, consensus should mitigate conflicts of interest. The tech industry is broad with competing interests---while we've discovered that we can't trust when a single company says, "Trust us with your data," we might be able to trust a diverse group of experts who say, "Trust us to ensure your privacy is respected. " Between this and government oversight, we can have a flexible and responsive regulatory authority that minimizes internal abuse.

\subsubsection{Engage engineers and designers to incorporate privacy by design}

Privacy must be a first-class requirement. However, we also need to realize that privacy is never going to be a one-size-first-all scenario. A possible approach would be to begin with the most restrictive privacy settings, then allow users to modify the settings when they take actions that may benefit from relaxing the settings.

Privacy is as important as performance and usability. During the development process, it should be considered a requirement, not as an optional feature.

\subsubsection{Empower users to understand the risks and benefits of immersive technology}

The foundations of US privacy law turn on the idea of 'reasonable expectation of privacy,' which is determined by society. Unfortunately privacy is a complex, poorly defined topic. We should help consumers understand the risks, not so that they each individually have to manually manage their privacy, but so they can advocate for themselves as a group.

Immersive technologies have amazing potential. They're already used to help reduce pain in burn patients \cite{hoffman2001effectiveness} and assist with complex manufacturing and maintenance \cite{palmarini2018systematic}. They give us new ways to experience the world around us and connect to our friends and families. It's unacceptable to say that using MR technology means that you have to sacrifice your privacy. We also don't want to lose out on the benefits by shunning MR because of privacy implications.

\subsubsection{Incorporate experts from other fields who have addressed similar problems}

Obviously, technology isn't the only field that needs ethics. Other fields have struggled with privacy, and we can learn from them. First, we need to accept the power that we, as technologists, have: we've built communication platforms; we've created robot vacuums; we've changed how news is disseminated; we've moved shopping and banking to the internet; we've digitized health records.

Now it's time for us to include medical ethicists, sociologists, anthropologists, economists, etc. to help figure out how we can address the ethical concerns discussed here and identify new issues.

Ethics are ideals---we will fall short of them. As long as we continue to work towards them, we can create a better future, one that is more respectful autonomy, self-expression, and privacy regardless of race of socioeconomic status. Taken in isolation, these steps likely won't accomplish this. They're complementary. By combining them, we can shape a better immersive future.

\section{Conclusions}
Mixed reality technologies are in the process of emerging into the mainstream. Although billions of dollars have been spent in development, they aren't yet entrenched in our lives. The technology is still malleable. By combining anticipatory foresight and reasoning with insights from similar entrenched technologies, like AI, this paper plausibly identifies and evaluates ethical issues in the field. MR occupies a complex space with significant overlaps and intersections with fields ranging from machine learning to healthcare and education; however, this paper identifies and discusses unique aspects of MR technology that set this area apart. This paper also recommends actions to take to mitigate negative impacts of the technology and enable not just ethical frameworks and guidance, but also ethical practice.

\balance
\bibliography{bib}{}
\bibliographystyle{plain}

\end{document}